\documentstyle[12pt,epsf]{article}
\newcommand{\beq}{\begin{equation}}
\newcommand{\eeq}{\end{equation}}
\newcommand{\beqa}{\begin{eqnarray}}
\newcommand{\eeqa}{\end{eqnarray}}

\begin{document}

\hfill KFA-IKP(TH)-1997-17


\hfill hep-ph/9709nnn

\vspace{1in}

\begin{center}

{{\Large\bf Isospin violation  
   in pion--nucleon scattering}}\footnote{Work supported
    in part by funds provided by the Graduiertenkolleg "Die Erforschung 
    subnuklearer Strukturen der Materie" at Bonn University.}

\end{center}

\vspace{.3in}

\begin{center}
{\large 
Ulf-G. Mei{\ss}ner$^\ddagger$\footnote{email: Ulf-G.Meissner@fz-juelich.de},
S. Steininger$^\dagger$$^\ddagger$\footnote{email: 
S.Steininger@fz-juelich.de}}

\bigskip

$^\ddagger${\it Forschungszentrum J\"ulich, Institut f\"ur Kernphysik 
(Theorie)\\ D-52425 J\"ulich, Germany}

\bigskip

$^\dagger${\it Universit\"at Bonn, Institut f{\"u}r Theoretische Kernphysik\\
Nussallee 14-16, D-53115 Bonn, Germany}\\

\bigskip

\end{center}

\vspace{.9in}

\thispagestyle{empty} 

\begin{abstract}\noindent
We construct the complete effective chiral pion--nucleon Lagrangian 
in the presence of virtual photons to one loop.
As an application, we consider  strong and
electromagnetic isospin violation for scattering of neutral pions off
nucleons. We show that for the  scattering lengths
these isospin violating terms are of the same size as the purely hadronic ones.
We also analyze isospin--violating effects for the
$\sigma$--term. These can be as large as 10\% for the absolute value
but are negligible for the shift to the Cheng--Dashen point. 
\end{abstract}

\vfill

\pagebreak

\noindent {\bf 1.} In his seminal paper in 1977, Weinberg pointed out
that reactions involving nucleons and {\it neutral} pions might lead
to gross violations of isospin symmetry~\cite{weinmass}. In particular,
he argued that the mass difference of the up and down quarks can
produce a 30\% effect in the difference of the $\pi^0 p$ and $\pi^0 n$
S--wave scattering lengths while these would be equal in case of
isospin conservation. This was later reformulated in more modern
terminology~\cite{weinmit}.
To arrive at the abovementioned result, Weinberg considered
Born terms and the dimension two symmetry breakers, one related to the
sum of the quark masses (i.e. the so--called $\sigma$--term) and the
other proportional to $m_u - m_d$. Bernard et al.~\cite{bkma} showed
that at this order there are other isospin--conserving terms which
make a precise prediction for the individual $\pi^0 p$ or $\pi^0 n$
scattering length very difficult. Nevertheless, Bernstein proposed to
measure $a(\pi^0 p)$ in neutral pion photoproduction off
protons~\cite{aron} based on the observation that the phase of the
electric dipole amplitude below the secondary $\pi^+ n$ threshold is 
related to $a(\pi^0 p)$ by unitarity (Fermi--Watson
theorem).\footnote{This idea was also mentioned by J.D. Bjorken, 
see~\cite{weinmass}.} With the
advent of high--precision data on pionic hydrogen and deuterium at 
PSI~\cite{sigg} and neutral pion photoproduction data from
MAMI~\cite{fuchs} and SAL~\cite{berg}, novel interest has been spurred
in separating electromagnetic effects and trying to filter out isospin
violating contributions. In addition, in the framework of some models
it has been claimed that the presently available pion--nucleon data
basis exhibits isospin violation of the order of a few 
percent~\cite{gibbs}\cite{mats}.\footnote{A status report with
emphasis on pion photoproduction is given in
the talk~\cite{ulfosaka}.} However, to really pin down isospin
breaking due to the light quark mass difference, one needs a machinery
that allows to {\it simultaneously} treat the electromagnetic and the
strong contributions. The only framework known at present allowing to
do just that is (baryon)
chiral perturbation theory. It is based on a consistent power counting
scheme which allows to order the various terms according to the number
of derivatives and/or meson mass insertions. This defines the
so--called chiral dimension corresponding to the number
of derivatives and/or pion mass insertions, with the pertinent
small parameters collectively denoted by $q$. For the pion--nucleon
system, the leading term is of dimension one and loops start to
contribute at two orders higher (in the heavy fermion
formalism~\cite{jm} which we will use). In what follows, we will add
to the well--known chiral effective pion--nucleon Lagrangian the terms
including virtual photons to one loop order. For that, we
construct the generating functional of two flavor heavy baryon chiral 
perturbation theory (HBCHPT)~\cite{bkkm} in the presence of virtual photons
and perform the necessary
renormalization. We also construct all finite terms up-to-and-including
third order in the chiral dimension. Throughout, we assign to the
electric charge a dimension one, based on the observation that
$e^2 \sim M_\pi^2/(4\pi F_\pi)^2 \sim 1/10 $ 
(with $M_\pi$ and $F_\pi$ the pion mass and decay constant, respectively).
We remark that some of these terms have also been constructed
by van Kolck based on a different power counting scheme~\cite{bira}  
and that Lebed and Luty have considered the SU(3) electromagnetic terms
relevant to the analysis of the baryon masses~\cite{lele}. 
We then work out the scattering lengths $a(\pi^0 p)$ and $a(\pi^0 n)$
to third order in small momenta, generalizing Weinberg's result and
discuss the relative size of the isospin conserving and violating terms.
We also consider the much discussed pion--nucleon $\sigma$--term and
analyze the isospin--violating contributions.
In this letter, we only work to third order in small momenta. As has
become clear from  studies concerning a broad variety of processes in
the pion--nucleon system, one has to go to fourth order to achieve
a high precision. The results presented here should therefore be
considered indicative and need to be supplemented by higher order
calculations.
 
\medskip

\noindent {\bf 2.} 
To introduce virtual photons in the effective pion--nucleon field
theory, consider first the nucleon charge matrix $Q = e \,{\rm diag}(1,0)$.
We form the matrices
\beq
Q_{\pm} = \frac{1}{2} \, (u \, Q \,u^\dagger \pm u^\dagger \, Q \, u) \,\, ,
\quad
\hat{Q}_{\pm} = Q_\pm - {1 \over 2} \langle Q_\pm \rangle \,\, ,
\eeq
where $U(x) = u^2(x)$ collects the pions and $\langle \ldots \rangle$
denotes the trace in flavor space. By construction, the
$\hat{Q}_{\pm}$ are traceless. Under chiral SU(2)$_L\times$SU(2)$_R$
symmetry, the $Q_\pm$ transform as 
\beq
Q_\pm  \to K \, Q_\pm  \, K^\dagger \quad ,
\eeq 
with $K$ the
compensator field representing an element of the conserved subgroup
SU(2)$_V$. Furthermore, under parity ($P$) and charge conjugation ($C$)
transformations, one finds
\beq
P \, Q_\pm \, P^{-1} = \pm \, Q_\pm \,\,\, , \quad 
C \, Q_\pm \, C^{-1} = \pm \, Q_\pm^T \,\,\, , 
\eeq
where $Q^T$ is the transposed of the matrix $Q$. For physical
processes, only quadratic combinations of the charge matrix $Q$
(or, equivalently, of the matrices $Q_\pm$) can appear (see also
ref.~\cite{mms}). It is now
straightforward to implement the (virtual) photons given in terms
of the gauge field $A_\mu$ in the effective pion--nucleon Lagrangian.
Starting from the relativistic theory and decomposing the spinor
fields into light (denoted $N$) and heavy components (velocity
eigenstates), one can
use path integral methods to integrate out the heavy components in a
systematic fashion. The effective Lagrangian decomposes into a sum
of terms with increasing chiral dimension (throughout, we use the
notation and conventions of the review~\cite{bkmrev}),
\beq
{\cal L}_{\pi N} = \sum_{n\ge 1} \, {\cal L}_{\pi N}^{(n)} 
= {\cal L}_{\pi N}^{(1)} + {\cal L}_{\pi N}^{(2)} + {\cal L}_{\pi N}^{(3)}
+ \ldots \,\,\, .
\eeq
In particular, to lowest order ($n=1$)
\beq
{\cal L}_{\pi N}^{(1)} = \bar{N} \,\biggl( i v \cdot \tilde{D} +
g_A \, S \cdot \tilde{u} \, \biggr) \, N\,\,\, , 
\eeq
with 
\beq \label{covder}
\tilde{D}_\mu = D_\mu -i \, Q_+ \, A_\mu \,\,\, , \quad 
\tilde{u}_\mu = u_\mu - 2 \, Q_- \, A_\mu \,\,\, ,
\eeq
and $v_\mu$ denoting the nucleons' four--velocity, $D_\mu =\partial_\mu
+ \Gamma_\mu$ the chiral covariant derivative, $S_\mu$ the
covariant spin--vector \`a la Pauli--Lubanski, $u_\mu = i u^\dagger
\nabla_\mu U u^\dagger$ and $g_A$ the axial--vector coupling 
constant. Here 
\beq
\nabla_\mu U = \partial_\mu U - i(v_\mu+a_\mu+QA_\mu)U
 +iU(v_\mu-a_\mu+QA_\mu) \,\, , 
\eeq
is the generalized pion covariant derivative containing the external
vector ($v_\mu$) and axial--vector ($a_\mu$)  sources. Note that to leading
order, no symmetry breaking appears in the effective pion--nucleon
field theory. At next order ($n=2)$, we have (remember that we count
$e$ as a small momentum, ${\cal O}(e) \sim {\cal O}(q)$) 
\beqa 
{\cal L}_{\pi N}^{(2)} &=& {\cal L}_{\pi N, {\rm str}}^{(2)} + 
{\cal L}_{\pi N, {\rm em}}^{(2)} \,\,\, , \\
{\cal L}_{\pi N, {\rm str}}^{(2)} &=&
\bar {N} \, \biggl\{ \frac{1}{2m}  ( v\cdot D)^2-\frac{1}{2m}{D \cdot D}  
+ c_1 \, \langle \chi_+ \rangle 
+ \biggl(c_2-\frac{{g}_A^2}{8{m}}\biggr)(v \cdot u)^2 \nonumber \\
&& \qquad + \,  c_3 \, u \cdot u + \ldots 
+  c_5 \biggl( \chi_+ - \frac{1}{2}\langle\chi_+\rangle \biggr) 
+ \ldots \biggr\} {N}  \,\, , \\
{\cal L}_{\pi N, {\rm em}}^{(2)} &=& \bar{N} \, F_\pi^2 \, \biggl\{ {f_1} \,
\langle Q^2_+ - Q^2_- \rangle + {f_2} \, \hat{Q}_+ \langle Q_+ \rangle
+ {f_3} \, \langle Q^2_+ + Q^2_- \rangle +
{f_4} \ \langle Q_+\rangle^2 \, \biggr\} \, N \,\,\, ,
\eeqa
with $\chi_+ = u^\dagger \chi u^\dagger + u \chi^\dagger u$ and
$\chi = 2B_0(s+ip)$ subsumes the external scalar and pseudoscalar sources. 
The external scalar source contains the quark mass matrix, $s(x) = 
{\rm diag}(m_u , m_d) + \ldots$ and $B_0 = |\langle 0| \bar q q |0\rangle
/ F_\pi^2$ measures the strength of the spontaneous symmetry breaking.
We assume $B_0 \gg F_\pi$ (standard CHPT).
Various remarks are in order. For the strong part, we have only
displayed the terms of relevance to the calculations discussed later.
In particular, the term $\sim c_5$ is the only one which leads to
an effect of the order $m_d - m_u$. This is exactly the B--term in
Weinberg's notation~\cite{weinmit}. All the others are isospin
conserving. The values of the low--energy constants (LECs) have been
determined in ref.~\cite{bkmlec}.
For the electromagnetic terms, we have written down the
minimal number allowed by all symmetries. Note that the last two terms
in ${\cal L}_{\pi N, {\rm em}}^{(2)}$ are proportional to $e^2 \bar{N}
N$. This means that they
only contribute to the electromagnetic nucleon mass in the chiral
limit and are thus not directly observable. However, this implies that
in the chiral two--flavor limit ($m_u =m_d=0$, $m_s$ fixed), the
proton is heavier than the neutron since it acquires an
electromagnetic mass shift. Only in pure QCD ($e^2=0$), this 
chiral limit mass is the same for the two particles.
Since we work to order
$q^3$ in the chiral expansion, we can always use the physical nucleon
mass, denoted $m$, and do not need to bother about its precise value
in the chiral limit. The numerical values of the electromagnetic LECs
$f_1$ and $f_2$ will be discussed below. The normalization factor of
$F_\pi^2$ in the electromagnetic pion--nucleon Lagrangian is
introduced so that the $f_i$ have the same dimension as the strong
LECs $c_i$. The corresponding dimension
two meson Lagrangian with virtual photons has been constructed 
in~\cite{egpdr}\cite{mms},
\beq \label{L2meson}
{\cal L}^{(2)}_{\pi\pi} = - {1\over 4}F_{\mu\nu}F^{\mu\nu} - {\lambda \over 2} 
(\partial_\mu A^\mu )^2 + {F_\pi^2\over 4} \langle \nabla_\mu U \nabla^\mu
 U^\dagger + \chi U^\dagger + \chi^\dagger U \rangle + C \langle Q U
 Q U^\dagger \rangle \,\, , 
\eeq
with $F_{\mu\nu} = \partial_\mu A_\nu - \partial_\nu A_\mu$ the photon field
strength tensor and $\lambda$ the gauge--fixing parameter (from here on,
we work in the Lorentz gauge $\lambda = 1$). It is important to stress
that in ref.~\cite{mms}, $Q$ denotes the {\it quark} charge
matrix. To make use of the {\it nucleon} charge matrix used here, we
perform a transformation of the type $Q \to Q + \alpha \, \bf 1$, with
$\alpha$ a real parameter and observe that $d \langle Q U Q U^\dagger \rangle
/ d\alpha \sim e^2 \, \bf 1$, i.e. to this order the difference between
the two charge matrices can completely be absorbed in an unobservable
constant term. To use the higher order terms constructed
in~\cite{mms}, one would have to rewrite them in terms of the nucleon 
charge matrix. In this paper, however, we do not need these terms and
thus do not consider them any further. Throughout, we work in the
$\sigma$--model gauge for the pions. In that case, the last term in
${\cal L}^{(2)}_{\pi\pi}$ leads only to a term quadratic in pion
fields. Consequently, the LEC $C$ can be
calculated from the neutral to charged pion mass difference since this
term leads to $(\delta M^2_\pi)_{\rm em} = 2e^2C/F_\pi^2$. This gives
$C = 5.9\cdot 10^{-5}\,$GeV$^{4}$.

\medskip

\noindent Consider now the neutron-proton mass difference. It is given
by a strong insertion $\sim c_5$ and an electromagnetic insertion
$\sim f_2$,
\beq \label{delnp}
m_n - m_p = (m_n - m_p)_{\rm str} + (m_n - m_p)_{\rm em} = 
4 \,c_5\, B_0\, (m_u - m_d) + 2 \, e^2 \, F_\pi^2 \, f_2  
+{\cal O}(q^4) \,\,\, .
\eeq
Note that one--loop corrections to the mass difference
only start beyond the order we are
considering here. This can be traced back to the fact that the
photonic self--energy diagram of the proton at order $q^3$ vanishes
since it is proportional to $\int d^dk \, [k^2 \, v\cdot k]^{-1}$
and the hadronic ${\cal O}(q^3)$ mass corrections are the same
for the $p$ and the $n$. To the
order we are working, the  electromagnetic LEC $f_2$ can therefore be
fixed from the electromagnetic proton mass shift, $ (m_n - m_p)_{\rm
em} = -(0.7\pm 0.3)\,$MeV, i.e. $f_2 = -(0.45\pm 0.19)\,$GeV$^{-1}$.
The strong contribution has been used in~\cite{bkmlec} to fix the LEC
$c_5 = -0.09 \pm 0.01\,$GeV$^{-1}$. Note that it is known that
one--loop graphs with an insertion $\sim m_d-m_u$ on the internal 
nucleon line, which in our counting appear at fourth order, can
contribute sizeable to the strong neutron-proton mass
difference~\cite{gl82}. Such effects go, however, beyond the accuracy
considered here but it underlines the need for a complete ${\cal
  O}(q^4)$ calculation.

\medskip

\noindent {\bf 3.} To go beyond tree level, we have to construct the
terms of order $q^3$. From the building blocks at our disposal, we can
construct a minimal set of
20 independent terms which are compatible with all
symmetries, following ref.~\cite{krause}. The number of possible terms
is limited due to the fact that at least two charge matrices must
appear. In particular, charge
conjugation allows to sort out terms which would otherwise be allowed.
Some of these terms are
accompanied by finite LECs where as the others absorb the divergences
appearing at one loop with LECs that are only finite after
renormalization. We now calculate these divergences. For doing that, we 
construct the
generating functional and extract the pertinent divergences making use
of heat kernel techniques for elliptic Euclidean differential
operators. We follow here the approach outlined in \cite{ecker}.
However, due to the presence of the virtual photons, we have
to expand the generating functional to second order in fluctuations 
around the classical solutions for the meson {\it and} photons fields, $U =
U^{\rm cl} + i u \, \xi\, u/F_\pi + \ldots$ and $A_\mu = A_\mu^{\rm cl} +
 \epsilon_\mu$, respectively. We consider only the the irreducible 
tadpole ($\Sigma_1$) and self--energy ($\Sigma_2$)  graphs,
leading to the one--loop functional at ${\cal O}(q^3)$,
\beqa
{\cal Z}_{\rm irr}[s,p,v_\mu,a_\mu;R_v;A_\mu] &=& \int d^4x \, d^4x'
\, d^4y \, d^4y' \, \bar{R}_v (x)  \,A_{(1)}^{-1} (x,y) 
\nonumber \\ &\times &  [ \Sigma_1 (y,y')
\delta^{(4)} (y-y') + \Sigma_2 (y,y')] \,A_{(1)}^{-1} (x',y') \,R_v (x') 
\eeqa
where $A_{(1)}^{-1}$ is the propagator for the nucleon field $N$ in
the presence of external fields and $R_v$ is the corresponding
velocity--projected nucleon source field. The extraction of the
$\Sigma_i$ in Euclidean space is straightforward. Consider first the
tadpole graph:
\beqa
\Sigma_1 &=& {i \over 8} \tau_k \biggl[ v \cdot \Sigma \, {\cal G} - 
{\cal G} \, 
v \cdot \stackrel{\leftarrow}{\Sigma} \biggr]_{ki} \, \tau_i - {3 \over 8} 
[\tau_i , Q_-] \, v_\mu \, {\cal G}_{i\mu} 
\nonumber \\
&+& {g_A \over 8}[\tau_i , [ \tau_j , S \cdot u ]] \, {\cal G}_{ij}
+ i g_A [\tau_i , Q_+ ] \, S_\mu \, {\cal G}_{i\mu} \,\,\, ,
\eeqa
with ${\cal G}$ the combined full pion ($G_{ij}$) and photon ($G_{\mu\nu}$)
propagator,
\beqa \label{piphopro}
{\cal G} = [ -\Sigma_\mu \Sigma_\mu + \Lambda]^{-1} =
\left(
\matrix { G_{ij}    &  G_{i\mu}   \nonumber \\
          G_{\nu j} &  G_{\mu\nu} \nonumber \\} 
\!\!\!\!\!\!\!\!\!\!\!\!\!\!\! \right) \, \, \, \, \, , 
\eeqa
where $i,j=1,2,3$; $\mu , \nu = 0,1,2,3$. $\Sigma_\mu$ is the combined
photon--pion covariant derivative,
\beq
\Sigma_\mu = \partial_\mu \, {\bf 1} + Y_\mu \,\,\,\, ,
\eeq
with
\beqa
Y_\mu^{AB} = \left(
\matrix{\displaystyle -{1\over 2}\langle [\tau_i , \tau_j ] \Gamma_\mu\rangle
        &\displaystyle  {F_\pi \over 2} \langle Q_- \tau_i \rangle
        \delta_{\mu\beta} \nonumber \\[0.2em]  \displaystyle
           -{F_\pi \over 2} \langle Q_- \tau_j \rangle \delta_{\mu\alpha}
        &\displaystyle 0 \nonumber \\}
\!\!\!\!\!\!\!\!\!\!\!\!\!\!\! \right) \, \, \, \, \, , 
\eeqa
where $A = (i,\alpha)$, $B = (j, \beta)$ and $\Gamma_\mu$ is the chiral
connection contained in the covariant derivative $D_\mu$,
Eq.(\ref{covder}). Furthermore, $\Lambda$ is defined via
\beqa
\Lambda^{AB} = \left(
\matrix { \sigma_{ij}       &  \gamma_{i\beta}   \nonumber \\
          \gamma_{\alpha j} &  \rho_{\alpha\beta} \nonumber \\} 
\!\!\!\!\!\!\!\!\!\!\!\!\!\!\! \right) \, \, \, \, \, , 
\eeqa
with
\beqa
\sigma_{ij} &=& {1\over 8} \langle \{ \tau_i , \tau_j \} \chi_+
\rangle - C\,F_\pi^2 \, \langle [\tau_i,Q_+] [\tau_j,Q_+] - 
[\tau_i,Q_-] [\tau_j,Q_-] \rangle \nonumber \\
&+& {1\over 8}\langle [ \tau_i , u_\mu ] [ \tau_j , u_\mu ] \rangle
- F_\pi^2 \, \langle Q_- \tau_i \rangle  \langle Q_- \tau_j \rangle
\nonumber \\
\rho_{\alpha\beta} &=& {3\over 2} F_\pi^2 \langle Q_-^2\rangle \, 
\delta_{\alpha\beta} \,\,\, , \quad \gamma_{\alpha i} = -{F_\pi \over
  2} \biggl[ i \langle u_\alpha [ \tau_i, Q_+] \rangle + \langle
\tau_i \, [ D_\alpha , Q_-] \rangle \biggr] \,\,\, .
\eeqa
Note that $\Lambda$ and its diagonal submatrices are symmetric under the
interchange of the pertinent indices.
The off--diagonal matrix--elements in ${\cal G}$, Eq.(\ref{piphopro}),
describe (virtual)  pion--photon transitions which can occur in the presence of
outgoing pions or other external sources. Similarly, the self--energy
graph leads to
\beqa
\Sigma_2 &=& {1\over 16} \, [v\cdot u , \tau_i] \, A_{(1)}^{-1} \,
[v\cdot u , \tau_j] \, {\cal G}_{ij} - Q_+ \, A_{(1)}^{-1} \, Q_+ \,
v_\mu \, v_\nu \,{\cal G}_{\mu\nu} 
\nonumber \\
&-& 4g_A^2 \, Q_- S_\mu \,  A_{(1)}^{-1} \, S_\nu \, Q_- \, {\cal G}_{\mu\nu}
+2g_A \, [ Q_+ \, v_\mu \, A_{(1)}^{-1} \, Q_- \, S_\nu  + 
           Q_- \, S_\mu \, A_{(1)}^{-1} \, v_\nu \, Q_+ ] \,  {\cal G}_{\mu\nu}
\nonumber \\
&-& {i \over 4} \, [v\cdot u , \tau_i] \, A_{(1)}^{-1} \,
(Q_+ \, v_\nu - 2g_A \, Q_- \, S_\nu) \, {\cal G}_{i \nu} - 
{i \over 4} \, (Q_+ \, v_\mu - 2g_A \, Q_- \, S_\mu) \,A_{(1)}^{-1} \,
[v\cdot u , \tau_j] \, {\cal G}_{\mu j} \nonumber \\
&+& {i\over 4}g_A \,[v\cdot u , \tau_i] \, A_{(1)}^{-1} \,{\cal G}_{ij}\,
S \cdot \stackrel{\leftarrow}{\Sigma}_{jk} 
\,\tau_k  + {i\over 4}g_A \, \tau_k \,
S \cdot \Sigma_{ki} \, A_{(1)}^{-1} \, [v\cdot u , \tau_j]\, {\cal G}_{ij} 
\nonumber \\
&+& g_A \,  (Q_+ \, v_\mu - 2g_A \, Q_- \, S_\mu) \,
A_{(1)}^{-1} \,{\cal G}_{\mu j} \,
 S \cdot \stackrel{\leftarrow}{\Sigma}_{jk} \,  \tau_k
- g_A^2 \, \tau_k \, S \cdot {\Sigma}_{ki} \, A_{(1)}^{-1}
\, {\cal G}_{ij} 
\, S \cdot \stackrel{\leftarrow}{\Sigma}_{jl}  \, \tau_l
\nonumber \\
&+& g_A \, \tau_k \,
S \cdot \Sigma_{ki} \, A_{(1)}^{-1} \,(Q_+ \, v_\mu - 2g_A \, Q_- \,
S_\mu) \, {\cal G}_{i \mu}  \,\,\, .
\eeqa
The functionals $\Sigma_{1,2} (y,y')$ are divergent in the coincidence
limit $y \to y'$. Following Ecker~\cite{ecker}, the
divergences can be extracted in a straightforward manner.\footnote{As
an excellent check, we recover the results of ref.~\cite{ecker} when
switching off the virtual photons.} The electromagnetic part of the 
dimension three  Lagrangian takes the form (after combining all finite
terms with the ones obtained after renormalization)
\beq \label{L3em}
{\cal L}^{(3)}_{\pi N,{\rm em}} = \sum_{i=1}^{20} g_i \, \bar N \,
{\cal O}_i \, N \,\, , 
\eeq
with the ${\cal O}_i$ monomials in the fields of dimension three. The
low--energy constants $g_i$ absorb the divergences in the standard
manner,
\beqa \label{L3LECs}
g_i &=& \kappa_i \, L + g_i^r (\mu ) \, \, , \\
L &=& {\mu^{d-4}\over 16\pi^2} \biggl\{ {1 \over d-4} - {1\over 2}
 \biggl[ \ln(4\pi ) + \Gamma '(1) +1 \biggr] \biggr\} \,\, ,
\eeqa
with $\mu$ the scale of dimensional regularization and $d$ the number
of space--time dimensions. 
\begin{table}[t]\begin{center}
\begin{tabular}{|r|c|c|} \hline
i  &  ${\cal O}_i$  &          $\kappa_i$ \\ \hline
1  & $ Q_+ \, S \cdot u \, Q_+  $ & $ 0 $ \\
2  & $ Q_- \, S \cdot u \, Q_-  $ & $ 2\,g_A (1-g_A^2) $ \\
3  & $ \langle Q_+ \, S \cdot u\rangle \, Q_+  $ & $ 4\,Z\,g_A (1-g_A^2) $ \\
4  & $ \langle Q_- \, S \cdot u\rangle \, Q_-  $ & $ -4\,Z\,g_A (1-g_A^2) $ \\
5  & $ \langle Q_+ \, S \cdot u\rangle \, \langle Q_+\rangle   $ & $ -2\,g_A
(1+Z-Z\,g_A^2) $ \\
6  & $ \langle Q_+^2 - Q_-^2\rangle \, S \cdot u  $ & $ -1/2\,g_A (1-8Z) $ \\
7  & $ \langle Q_+^2 + Q_-^2\rangle \, S \cdot u  $ & $ -3/2\,g_A$ \\
8  & $ \langle Q_+\rangle ^2 \, S \cdot u  $ & $ -2 Z\,g_A $ \\
9  & $ \langle Q_- \, v \cdot u\rangle \, \langle Q_+\rangle   $ 
   & $ 1- 3\,g_A^2 $ \\
10 & $ \langle Q_- \, v \cdot u\rangle \, Q_+  $ & $  1+ 3\,g_A^2 $ \\
11 & $ \langle Q_+ \, v \cdot u\rangle \, Q_-  $ & $  -(1+ 3\,g_A^2) $ \\
12 & $ Q_+ \,v \cdot u \,Q_- + Q_- \, v \cdot u \,Q_+$ & $ -3(1- g_A^2)$ \\
13 & $ \langle Q_+\rangle \, Q_+ \, i v \cdot D + \mbox{h.c.}  $ & $ -2 $ \\
14 & $ \langle Q_+^2 - Q_-^2\rangle \, i v \cdot D  + \mbox{h.c.}  $ & $
-1/2-3/4\,g_A^2 (1+8Z) $ \\
15 & $ \langle Q_+^2 + Q_-^2\rangle \, i v \cdot D  + \mbox{h.c.}  $ & $ -1/2 +
3/4\,g_A^2 $ \\
16 & $ \langle Q_+\rangle ^2 \, i v \cdot D  + \mbox{h.c.} $ & $ 1+3Z\,g_A^2 $ \\
17 & $ [Q_+,i v \cdot c_+]  $ & $ -2 $ \\
18 & $ [Q_-,i v \cdot c_-]  $ & $ -1/2\,(1-9\,g_A^2) $ \\
19 & $ [Q_+,i S \cdot c_-]  $ & $ 0 $ \\
20 & $ [Q_-,i S \cdot c_+]  $ & $ 0 $ \\
\hline
\end{tabular}
\caption{Operators of dimension three and their $\beta$--functions.
Here, $Z = C/F_\pi^4$.}
\end{center}
\end{table}
\noindent The explicit expressions for the operators
${\cal O}_i$ and their $\beta$--functions $\kappa_i$ are collected in 
table~1. The $g_i^r (\mu)$ are the renormalized, finite and
scale--dependent low--energy constants. These can  be fixed by data or
have to be estimated with the help of some model. They obey the
standard renormalization group equation. To arrive at the
terms in the table, we have used the relation
\beqa 
[D_\mu , Q_\pm] &=& -{i \over 2} [u_\mu , Q_\mp ] + c_\mu^\pm \,\,\, 
\nonumber \\
c_\mu^\pm &=& {1\over 2} \biggl\{ u (\partial_\mu Q - i [v_\mu-a_\mu ,
Q]) u^\dagger \pm u^\dagger (\partial_\mu Q - i [v_\mu +a_\mu ,Q])u 
\biggr\} \,\,\,\, .
\eeqa
Furthermore, one could use the relation
\begin{equation}
[\nabla^\mu,u_\mu] = \frac{i}{2} \, \chi_- - \frac{i}{4} \,
\langle\chi_- \rangle + i\frac{4C}{F^2}\, [Q_+,Q_-]
+ {\cal O}(q^4) \,\, ,
\end{equation}
to rewrite some of the terms tabulated. In fact, this has been done by
Ecker~\cite{ecker} for the case without virtual photons ($e^2=0$).
We prefer not to do this and therefore Ecker's operator ${\cal
  O}_8^{\rm str} =[\chi_-,v \cdot u]$ in our basis reads ${\cal O}_8^{\rm str} 
= -2 [ [i\nabla^\mu,u_\mu], v \cdot u]$. 
A few remarks concerning the operators given in table~1 are in order.
First, ${\cal O}_7$ and ${\cal O}_8$ only lead to an electromagnetic
renormalization of $g_A$ and their effects can thus completely be absorbed
in the physical value of the axial--vector coupling constant. The operators
${\cal O}_{17,\ldots,20}$ are only of relevance for processes with external
axial--vector fields (or low--energy manifestations of $Z^0$--exchange)
and ${\cal O}_{13,\ldots,16}$ can be eliminated by
use of the nucleon equations of motion, i.e. they are not relevant for
reactions involving on--shell nucleons.
The general effective pion--nucleon Lagrangian with virtual
photons constructed here allows now to systematically investigate the
influence of isospin--breaking due to the quark mass difference $m_u -
m_d$ and the dual effects from electromagnetism. Of particular
importance are the processes $\pi N \to \pi N$, $\gamma N \to \pi N$
and $\pi N \to \pi \pi N$ since a large body of precise low--energy
data exists which can be analysed within the framework outlined. Here,
we concentrate on one particular reaction, namely elastic neutral pion
scattering off nucleons.

\medskip

\noindent {\bf 4.} We are now in the position to evaluate the isospin
violating corrections to the $\pi^0 p$ scattering length. To one--loop
order, we consider the isospin--conserving (IC) tree and loop graphs
already evaluated in~\cite{bkma}, the isospin--violating strong tree graph
$\sim m_u -m_d$~\cite{weinmass} (cf. fig.1a) and the
isospin--violating loop graphs (cf. fig.1b,c). For the latter two,
isospin violation comes in solely through the charged to neutral pion
mass difference, as indicated by an insertion $\sim C$ on the
internal pion lines. The rescattering type of diagram (fig.1b) is well
known from neutral pion photoproduction, where it leads to the cusp
effect in the electric dipole amplitude~\cite{bkmzpc}. 

\begin{figure}[t]
   \vspace{0.5cm}
   \epsfxsize=11.9cm
   \centerline{\epsffile{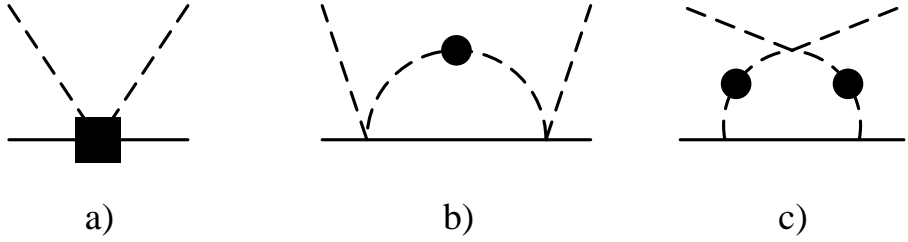}}
   \vspace{0.2cm}
   \centerline{\parbox{15cm}{\caption{\label{fig1}
   Graphs contributing to isospin violation in  $\pi^0$--proton 
   scattering. Solid and dashed lines denote nucleons and pions, in
   order. The heavy dot and the box refer to the 
   em counterterm at order $e^2$, i.e. the term proportional to $C$,
   and the dimension two strong insertion $\sim m_u -m_d$, respectively. 
   Diagram a) has previously been considered by 
   Weinberg~{\protect \cite{weinmass}}.  
}}}
\end{figure}

\noindent Evaluating all diagrams, the resulting expression reads
\beq
a(\pi^0 p) = {1\over 4\pi} \biggl( 1 + {M_{\pi^0}\over m_p}
\biggr)^{-1} \, \biggl( a_{\rm IC}^{(2)} +  a_{\rm IV}^{(2)} +
a_{\rm IC}^{(3)} +   a_{\rm em}^{(3)} \biggr) \,\,\, ,
\eeq
with
\beqa
 a_{\rm IC}^{(2)} &=& {2M_{\pi^0}^2 \over F_\pi^2} \biggl( c_2 + c_3
-2c_1 - {g_A^2\over 8m} \biggr) \,\,\,\, , \\
 a_{\rm IV}^{(2)} &=& -{2 \, c_5 \over F_\pi^2} \, B_0 \, (m_u-m_d)
\,\,\,\, , \\
 a_{\rm IC}^{(3)} &=& { 3 \, g_A^2 \,M_{\pi^0}^3 \over 64 \, \pi \,
   F_\pi^4} \,\,\,\, , \\ \label{rescatt}
 a_{\rm em}^{(3)} &=& 
- {M_{\pi^0}^2 \over 8 \pi F_\pi^4} \, 
\sqrt{M_{\pi^+}^2 - M_{\pi^0}^2} +
{3 g_A^2 M_{\pi^0}^2 \over 32 \, \pi F_\pi^4}\,
(M_{\pi^+} - M_{\pi^0} ) 
\,\,\,\, .
\eeqa  
In Weinberg's analysis, only dimension two terms were
considered~\cite{weinmass}. He calculated the Born piece and the
$\sigma$--term, i.e. the contributions $\sim g_A^2$ and $\sim c_1$ in 
$a_{\rm IC}^{(2)}$ as well as  $a_{\rm IV}^{(2)}$. We recover, of
course, these particular terms once we account for the differences in
notation. For the neutron case, one finds a similar result. Most
interestingly, the difference between $a(\pi^0 p)$ and $a(\pi^0 n)$
is given {\it entirely} by the dimension two term $\sim m_u -m_d$,
thus Weinberg's result
\beq
a(\pi^0 p) - a(\pi^0 n) = {1\over 4\pi} \biggl( 1 + {M_{\pi^0}\over m_p}
\biggr)^{-1} \,\biggl(-{4\,c_5\over F_\pi^2} \, B_0 \, (m_u-m_d) \biggr) 
+ {\cal O}(q^4) \,\,\, ,
\eeq
is only affected at next-to-next-to-leading order, ${\cal O}(q^4)$.
Let us now analyse
the various pieces contributing to the neutral pion--proton scattering
length. Using as input $F_\pi = 92.4\,$MeV, $g_A = 1.29$ (from the
Goldberger--Treiman relation with $g_{\pi N} = 13.05$ which is
equivalent to using $g_A =1.26$ and explicitely keeping the dimension
three operator appearing in the Goldberger--Treiman discrepancy), $M_{\pi^0} 
= 134.97\,$MeV, $M_{\pi^+} = 139.57\,$MeV, $m_p = 938.27\,$MeV and the
best fit values from~\cite{bkmlec}, $c_1 = -0.91\,$GeV$^{-1}$,
$c_2 = 3.25\,$GeV$^{-1}$ and $c_3 = -5.16\,$GeV$^{-1}$ (see also the
discussion below), we find\footnote{A different determination
of these LECs leads to the same values, see~\cite{moj}.}
\beqa
 a_{\rm IC}^{(2)} &=& -1.33 \, {\rm GeV}^{-1} \,\,\, , \quad
 a_{\rm IV}^{(2)}  =  +0.12 \, {\rm GeV}^{-1} \,\,\, , \\
 a_{\rm IC}^{(3)} &=& +0.84 \, {\rm GeV}^{-1} \,\,\, , \quad
 a_{\rm em}^{(3)}  =  -0.30 \, {\rm GeV}^{-1} \,\,\,\, .
\eeqa
The total scattering length comes out to be $a(\pi^0 p) = -8.8 \cdot 
10^{-3}/M_\pi$. Notice the large cancellations between the
isospin--conserving terms of dimension two and three, which make a 
precise knowledge of the LECs $c_{1,2,3}$ necessary to sharpen this
prediction. For this set of the $c_i$, $a(\pi^0 p)$ decomposes as
\beq
a(\pi^0 p) = a(\pi^0 p)_{\rm str, IC} +  a(\pi^0 p)_{\rm str, IV} + 
a(\pi^0 p)_{\rm em} = (-4.8 - 1.1 - 2.9)  \cdot 10^{-3}/M_\pi \,\,\, ,
\eeq
which shows that the isospin--violating terms are of the same size as
the isospin conserving ones. In particular, the electromagnetic
correction is bigger than the one due to the quark mass difference and
it is  dominated by the rescattering graph (fig.~1b and first
term in Eq.(\ref{rescatt})). We also
find that $a(\pi^0 p) - a(\pi^0 n) = -2.2 \cdot 10^{-3}/M_\pi$ and
thus the strong isospin violation is 25\% in the difference. This
agrees with Weinberg's finding~\cite{weinmass}. His numbers are
different from ours because his isospin--conserving strong contribution
is not the same for the reasons mentioned above and also, he used
SU(3) arguments to estimate $c_5$, which leads to a somewhat larger
value than the one we use. However, the isospin violating effects
might even be bigger. In the world of perfect isospin symmetry,
the isospin--even scattering length $a^+$ is given exactly by the terms 
$a_{\rm IC}^{(2,3)}$. This hadronic contribution has recently been
measured at PSI by combining the strong interaction shift measurements
of pionic hydrogen and deuterium, $a^{+}=(0...5) \cdot 
{10^{-3}}{M_\pi^{-1}}$~\cite{sigg}.\footnote{Note that this 
  result might still change
  a bit since a more sophisticated treatment of
  Doppler--broadening for the width of the pionic hydrogen has to be 
  performed. Also, the PSI--ETHZ group did not yet quote a value for 
  $a^+$. We rather used their figure combining the published H and d 
  level shift results
  to get the band given. If one combines their numbers from the pionic
  hydrogen shift and width measurements, one gets a negative value for
  $a^+$. This agrees with a novel pionic hydrogen 
  measurement at PSI leading to $a^{+}=
  (-8...0)\cdot  {10^{-3}}{M_\pi^{-1}}$~\cite{schroeder}.} 
This number is, however, not consistent with $\pi$N partial wave
analysis, the value obtained in the SM95 solution of the VPI group is
$a^{+}=-3.0\cdot {10^{-3}}{M_\pi^{-1}}$~\cite{vpi} where as the
standard  Karlsruhe-Helsinki value~\cite{koch} is
$a^{+}=-(8.3 \pm 3.8) \cdot {10^{-3}}{M_\pi^{-1}}$. For that reason,
in the fit which determined the $c_i$, $a^+$ was allowed to vary in
the wide band between $-10 \ldots +10 
\cdot {10^{-3}}{M_\pi^{-1}}$~\cite{bkmlec}. A recent calculation
to third order in small momenta for pion scattering off deuterium,
based on the formalism developed in~\cite{wein1}, leads to
$a^+ = -(2.6\pm 0.5)\cdot {10^{-3}}{M_\pi^{-1}}$~\cite{bblm}.   

\medskip

\noindent Of course, one can and should extend these considerations
to elastic scattering of charged pions and charge exchange reactions.
However, for these processes the isospin--conserving terms are much
larger and also the effects of virtual photons are more dramatic. 
Nevertheless, in the light of the recent claims that the low--energy
pion--nucleon scattering data basis shows isospin violation of the
order of 5\%~\cite{gibbs}\cite{mats}, this issue has to be pursued.
The framework presented here should allow to perform these
calculations in a {\it systematic} fashion.

\medskip

\noindent {\bf 5.} Furthermore, the $\sigma$--term
analysis might also be strongly affected by isospin--violating
effects (as already stressed by Weinberg). Here, we  reconsider it to
third order in small momenta. The scalar form factor is defined as the
expectation value of the QCD quark mass term in a proton (neutron)
state (note that due to the isospin--breaking, we have to work in
the basis of physical states),
\beq
\sigma_p (t) = \langle p' | m_u \,\bar u u + m_d \,\bar d d | p\rangle~,
\eeq
with $|p\rangle$ a proton state of momentum $p$ and $t = (p'-p)^2$
the invariant momentum transfer squared. A similar definition applies
for the neutron. The precise way of how to calculate this matrix
element in HBCHPT is spelled out in ref.~\cite{bkkm}.
The $\sigma$--term is simply the scalar form factor
at $t=0$. Rewriting the quark mass term as a sum of an
isoscalar and isovector component, one recovers the standard
definition of the  $\sigma$--term in the limit $m_u = m_d = \hat m$.
To ${\cal O}(q^3)$, the scalar form factor receives contributions 
from dimension two tree level insertions $\sim c_1$ and $\sim c_5$
as well as from finite one--loop graphs. In these, isospin violation
comes in through the neutral to charged pion mass
difference. Alltogther,
we find 
\begin{eqnarray}
\sigma_p (t) & = & 
-4 M_{\pi^0}^2 \, c_1 - 2 B_0 (m_u-m_d) \, c_5
-\frac{g_A^2 \, M_{\pi^0}^2}{32\pi F_\pi^2} (2 M_{\pi^+}^2 + M_{\pi^0}^2)
\nonumber \\
& &
+\frac{g_A^2 \, M_{\pi^0}^2}{128\pi F_\pi^2} 
\left\{
2(t-2M_{\pi^+}^2) \int_0^1 dx \left[M_{\pi^+}^2+t x(x-1)\right]^{-1/2}
\right.  \nonumber \\
& &\left. \qquad\qquad\quad
+(t-2M_{\pi^0}^2) \int_0^1 dx \left[M_{\pi^0}^2+t x(x-1)\right]^{-1/2}
\right\} \,\,\,\, , \\
\sigma_n(t) &=& \sigma_p(t) + 4 B_0 \, (m_u-m_d) \, c_5 \quad .
\end{eqnarray}
The scalar form factor at the Cheng--Dashen point  $t =
  2M_{\pi^+}^2$ can be easily worked out from this using
\beq
\int_0^1 dx \left[M^2+t x(x-1)\right]^{-1/2} = {1\over \sqrt{t}} \ln 
{2M+\sqrt{t} \over 2M- \sqrt{t} } \quad .
\eeq
The difference $\sigma_p (2M_{\pi^+}^2) - \sigma_p (0)$ is entirely 
given by the one--loop
contributions since the tree insertions are momentum--independent.
For the numerical analysis, we use the same parameters as above. We
note that there is currently some controversy about the precise value
of the $\sigma$--term at the Cheng--Dashen point, which might change
some of the values quoted here. It is, however, straightforward to
update these once the controversy has been settled. We find
(for the isospin--conserving piece, we use the charged pion mass)
\beq
\sigma_p(0) = \sigma_p^{IC} (0) + \sigma_p^{IV} (0) 
= 47.2~{\rm MeV} - 3.9~{\rm MeV}  = 43.3~{\rm MeV} \,\,\, ,
\eeq
which means that the isospin--violating terms reduce the proton
$\sigma$--term by $\sim 8\%$. The electromagnetic effects are again
dominating the isospin violation since the strong contribution 
is just half of the strong
proton--neutron mass difference, 1~MeV. Furthermore, one gets
$\sigma_p (2M_{\pi^+}^2) - \sigma_p (0) = 7.5$~MeV, which 
differs from the result in the isospin limit (7.9~MeV) 
by 5\% and is by about a factor
two too small when compared to the dispersive analysis of ref.~\cite{gls}.
This small difference of 0.4~MeV
 is well within the uncertainties related to the
so--called remainder at the Cheng--Dashen point~\cite{bkmcd}.

\medskip

\noindent {\bf 6.} To summarize, we have constructed the complete
chiral effective pion--nucleon Lagrangian including the
effects of virtual photons to third order in small momenta (i.e.
in the one--loop approximation). Counting
the electric charge as a small momentum, there are in total 4 and 20 terms
contributing to the electromagnetic
 Lagrangian at dimension two and three, respectively
(some of these do not appear in observables, e.g. at dimension
two one has effectively two terms). As an application, we have
considered  isospin violation in  elastic $\pi^0$--nucleon scattering.
We have sharpened Weinberg's time--honoured
calculation~\cite{weinmass} by extending
it to third order in small momenta. In addition to the sizeable
isospin violation from the quark mass difference, there is an even
bigger effect due to virtual photons hidden in the pion mass
difference. Weinberg's prediction concerning the difference of the
S-wave scattering lengths $a(\pi^0 p) - a(\pi^0 n)$ is, 
however, not affected at third order. We have also considered
the scalar form factor and shown that the proton $\sigma$--term is
reduced by about 8\% due to isospin violation. The effect in the
difference $\sigma_p (2M_{\pi^+}^2) - \sigma_p (0)$ is well within
the theoretical uncertainties related to scalar meson exchange, strangeness
effects and so on (for details, see ref.~\cite{bkmcd}). 
In a forthcoming publication, we will present results also
for the channels involving charged pions  
and thus clarify the origin of the isospin violation claimed to be
seen in low--energy pion--nucleon scattering. Ultimately, these
calculations should be carried out at fourth order in small momenta
which will require some work.

\vskip 0.5cm

\section*{Acknowledgements}

We are grateful to J\"urg Gasser, Norbert Kaiser, Heiri Leutwyler and
Guido M\"uller for useful comments.

\vskip 2cm


\end{document}